# Characterisation of superfluid vortices in helium II

Gregory P. Bewley[1,2], Daniel P. Lathrop[1] & Katepalli R. Sreenivasan[3,1]

[1]*Department of Physics, Institute for Research in Electronics and Applied Physics, Institute for Physical Sciences and Technology, University of Maryland, College Park, MD, 20742, USA*

[2]*Department of Mechanical Engineering, Yale University, New Haven, CT, 06520, USA*

[3]*International Centre for Theoretical Physics, Trieste, Italy*

**Matter at low temperatures exhibits unusual properties such as superfluidity, superconductivity, Bose-Einstein condensation, and supersolidity[1]. These states display quantum mechanical behaviours at scales much larger than atomic dimensions. As in many phase transitions, defects can occur during the transition to the low temperature state. The study of these defects yields useful information about the nature of the transitions and of the macroscopic states themselves. When cooled below the lambda temperature (~2.172 K), liquid helium acquires superfluid[2] properties, and the defects in the superfluid take the form of line vortices with quantized circulation[3]. The formation of these vortices has been suggested as a model for cosmological structure formation[4,5]. Here we visualize these superfluid (or quantized) vortices by suspending micron-sized solid particles of hydrogen in the fluid. The particles at low concentrations arrange themselves with nearly equal spacing along vortex lines. The number density of the vortex lines in a steadily rotating state compares well with Feynman's prediction[6]. For the first time, it is possible to observe superfluid turbulence in which the superfluid vortices interact[7], and exist in networks with complex branching. Our method makes it possible to characterize the vortices during the transition across the lambda point and in the turbulent state itself.**

We have developed a method for producing micron-sized particles of solid hydrogen and have used this to visualize superfluid vortices in helium II (the lowest temperature state of liquid helium, which occurs below the lambda point[2]). We prepare the suspension of particles in helium I (liquid helium above the lambda point) and lower its temperature to a desired state below the lambda point where superfluid vortices are formed spontaneously. The hydrogen particles are only small if injected into helium I, rather than helium II.

Our basic observation is that when the helium is cooled below the lambda point, a fraction of the particles collects along lines (see Fig. 1). Most particles, however, remain randomly distributed. Both along the visible lines and in the bulk, the particles move through the illuminated volume, and it appears that the two motions are independent of each other. This is consistent with Landau's picture[8] of helium II as composed of two interpenetrating fluids, one normal and one superfluid. We establish below that the lines along which the particles are trapped are indeed the superfluid vortex lines.

As the fluid undergoes the phase transition across the lambda point, we have found that changes occur in four successive and distinct stages. At a fixed cooling rate, the temperature falls steadily before transition, then changes more slowly, and stalls for tens of seconds near transition. Subsequently, the temperature continues to fall, the vigorous convection from boiling ceases, and the fluid becomes relatively quiescent. The particles then collect on short lines of ~100 μm length. The concentration of lines is ~$10^6$ m$^{-2}$. The occurrence and growth of these features are shown in Fig. 2. Finally, after some minutes and a few hundredths of a degree below the lambda point, a state is reached in which extended thread-like features of about 5 mm length drift through the illuminated volume. The structures formed by the particles during transition appear to depend on the cooling rate and the kinetic energy initially present in the fluid. Because of our relatively slow cooling rate, the generation of defects in the phase transition and the decay of turbulence are comingled in a way not present in existing theory.

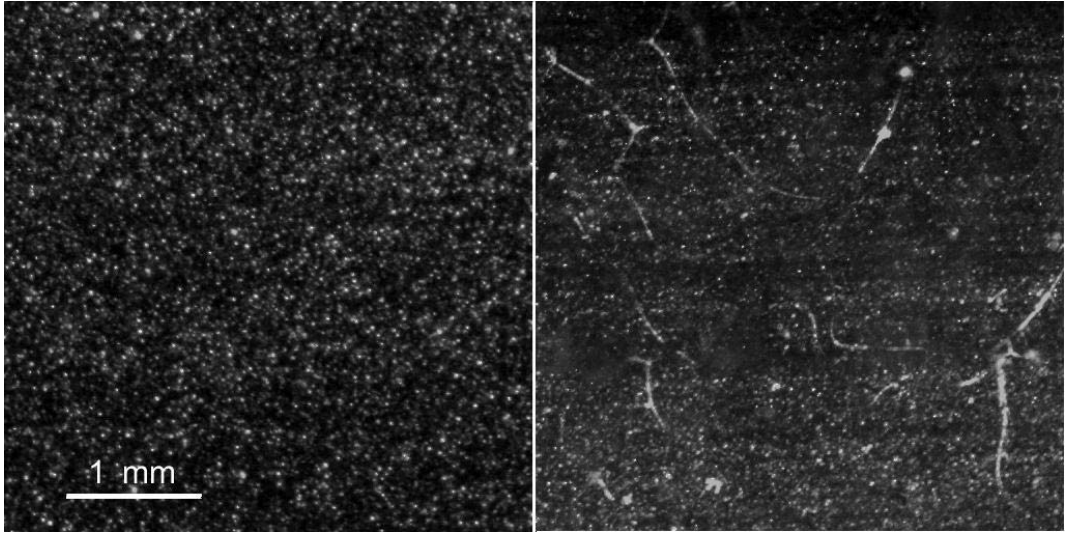

Figure 1: Images above and below the transition. For temperatures slightly above the lambda transition (left), hydrogen particles are randomly dispersed and make it possible to track the flow of the liquid helium. Below the transition (right), some fraction of the particles become trapped on lines in the flow. The particles are illuminated with a laser sheet; these are side views. Our data show that the lines thus observed are quantized superfluid vortices.

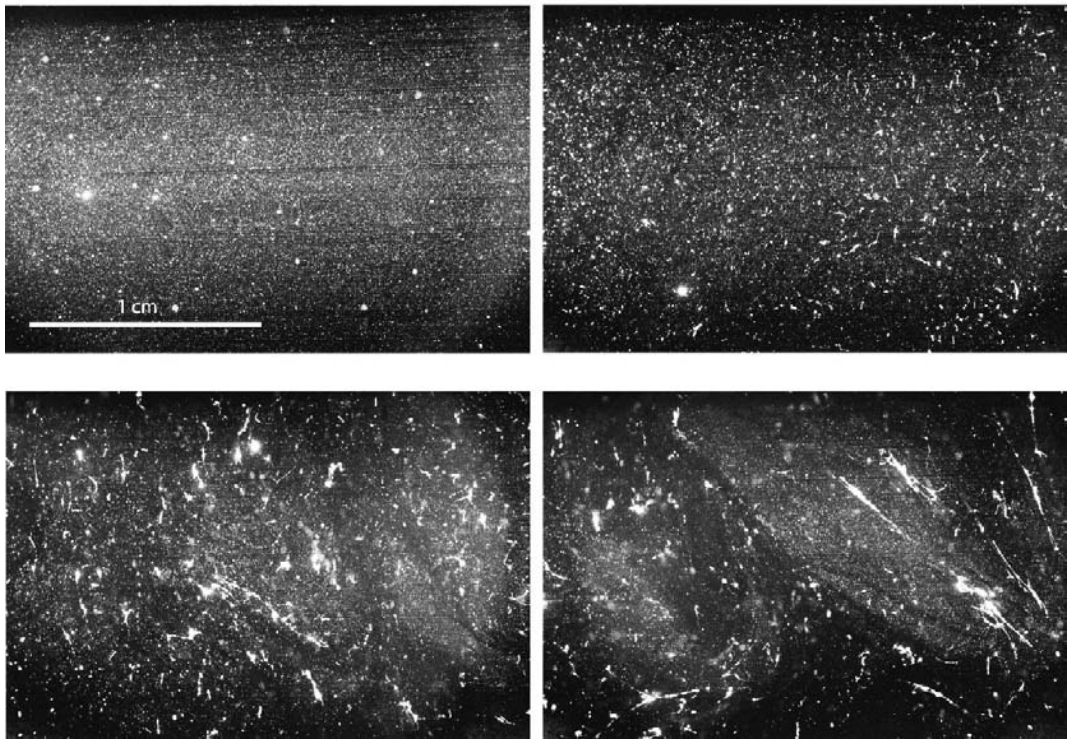

Figure 2: The development of vortices through the transition. The sequence of images as liquid helium is cooled below the lambda line shows the particles

becoming arrayed on a progressively coarser assemblage of vortices. These images were taken at temperatures and times after passing the lambda point given by (a) 2.172 K, 14 s; (b) 2.160 K, 37 s; (c) 2.148 K, 83 s; and (d) 2.125 K, 102 s.

Steady rotation of helium II produces a triangular array of quantized vortices[2], aligned with the axis of rotation. All the rotation of the superfluid resides in these vortices. The number density of the vortex lattice is proportional to the rotation rate, as predicted by Feynman's rule[6], $n_o = 2\Omega/\kappa \sim 2000\,\Omega$ lines/cm$^2$, where $\Omega$ is the angular velocity of the container in radians/s, $\kappa = h/m_4$ the quantum of circulation, $h$ Planck's constant and $m_4$ the mass of a helium atom. We rotate our Dewar to test whether the visible lines follow these expectations. Because many lattice defects appear at high particle concentration, due in part to particle interactions, we perform experiments at low particle concentration. We then observe that most of the visible lines in a steadily rotating state align parallel to the axis of rotation. The lines are also uniformly spaced, as shown in the Fig. 3. We measure the spacing of lines from such images and compute their number density under the assumption that our laser sheet illuminates a cross section of a triangular lattice, and is thin relative to the lattice spacing. The thickness of our laser sheet can lead to an over-count of the vortices, thus contributing to an apparently larger density, but the data plotted in Fig. 3 are consistent with Feynman's theory.

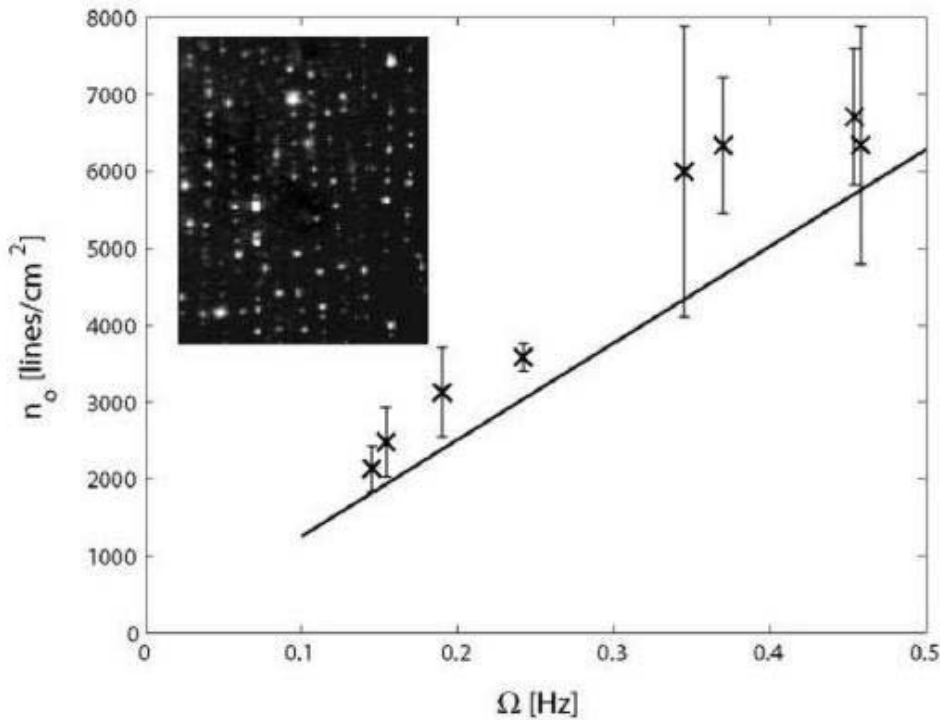

Figure 3: Line density in the rotating state. The line spacing of the superfluid vortices in rotating experiments, plotted here as a number density, is consistent with the Feynman rule for vortex density for solid body rotation. The error bars show the standard deviation of individual realizations. The inset shows an image with particles at low density aligned on vertical superfluid vortices.

We also observe a number of unexpected phenomena. At low particle densities in non-rotating states, we have frequently observed particles arrayed in what might be described as dotted lines, Fig. 4a. The striking uniformity of the spacing of particles appears not to be affected much by the particle size. In other cases, the observed lines show branching, typically with three lines joining at a node, Figs. 4b,c. When the rotating state is not in equilibrium the vortices can interact, showing wave motion[2], reconnection[9] and a fascinating type of superfluid turbulence[10]. We have also observed waves in the rotating vortices consistent with Tkachenko waves[2].

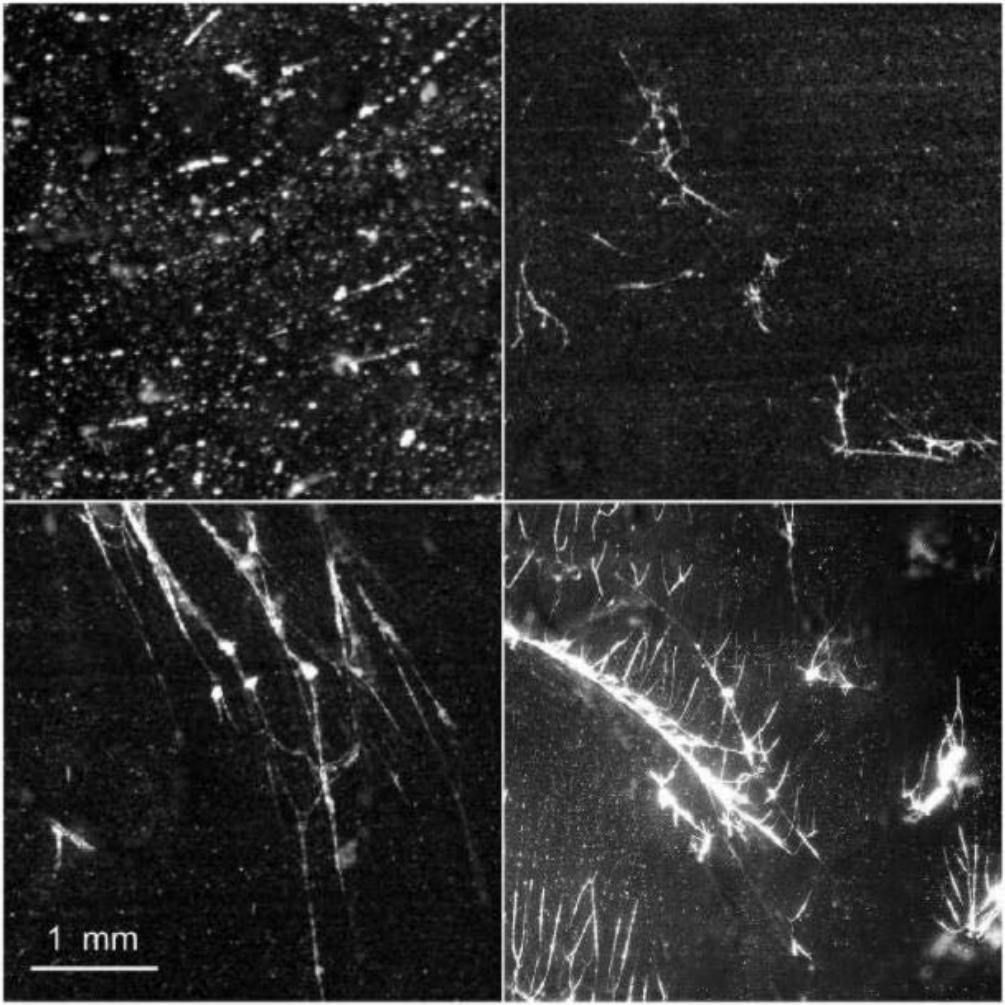

Figure 4: New phenomena observed in the superfluid state. These images show (a) particles evenly spaced on line vortices, (b) branching of lines, (c) superfluid vortex networks, and (d) superfluid vortex defects in the rotating state.

Other researchers have detected quantized vortices indirectly by a variety of means, including using the vortex attenuation of second sound in the superfluid[11]. Until now, individual vortices have been resolved only in a two dimensional projection in cases where the vortex lines were parallel to each other[12]. Packard's group observed vortices at their intersection with the top boundary of the fluid, and determined that in solid body rotation, the superfluid formed a triangular lattice of vortices. They found that the vortex area density agrees with Feynman's rule, although, like us, they too consistently measured higher densities. Our method is unique in allowing the study of the formation, dynamics, and statistical properties of superfluid vortices.

Although no comprehensive microscopic description of superfluid helium exists, several ideas and models help to interpret our observations. According to the two-fluid model[8], helium II can be regarded as being made up of two interpenetrating fluids, the superfluid and the normal fluid, with different density and velocity fields, $\rho_s$, $v_s$, $\rho_n$, and $v_n$, and the quantized vortices are defects in the superfluid component. The superfluid velocity in the bulk is a potential flow with no vorticity, $v_s = \kappa \nabla \phi$ and $\nabla \times v_s = 0$, except on the line defects where all the rotation is concentrated[3]. The superfluid component moves about these defects with a quantized circulation $\kappa$. The vortices are a close approximation to an ideal vortex line[13], with a locally cylindrical flow with the azimuthal velocity at a radius $r$ given by $v_\phi = \kappa/2\pi r$.

Macroscopic suspensions in the liquid are attracted to the core of quantum vortices by the steep pressure gradient supporting the circulating superfluid[14]. Whether or not a particle is trapped in a vortex core depends on its diameter and on the relative velocities and densities of the normal and superfluid components[15]. To make plausible estimates, we first note that a pressure field $P = -\rho_s \kappa^2/8\pi^2 r^2$ is set up, for $r$ larger than the core radius, by the superfluid rotation about the vortices. Presumably, the pressure flattens out for $r$ smaller than the core radius, leading to a pressure minimum around the edge of the vortex core. The force that a particle feels due to the gradient of this pressure is balanced by the Stokes drag[13] due to its motion relative to the normal fluid component. This balance yields $6\pi a \mu v_n = (4/3)\rho_s \pi a^3 \nabla P$, where $a$ is the radius of the particle and $\mu$ and $v$ are the viscosity and velocity of the normal fluid. The greatest attractive force will be felt toward a vortex line when the particle is just outside the core, and, if the Stokes drag exceeds this force, the particle will not be trapped. By using the particle radius as the distance from the core of maximum attractive force, we estimate the maximum allowable normal fluid velocity for trapping to occur as $v_n = \rho_s \kappa^2/18\pi^2 a \mu$. At 2.10 K, a 0.5 μm particle is dislodged by a relative velocity of the normal fluid of 2.5 mm/s and above. A big particle of the order of 5 μm diameter particle will be dislodged by a miniscule relative velocity of 0.25 mm/s. The absolute velocities of particles in our flows are typically 1 to 10 mm/s. These estimates point to the sensitivity of our technique to particle size, and the need for particles smaller than 1 μm in size for most cases of interest.

The effects of particles on vortex behaviour may explain some of our observations. For example, more than one vortex line may trap a particle, and reciprocally, particles may hold together collections of vortices. This could allow branches and bundles of vortices to form when it may otherwise have been energetically unfavourable. Stokes theorem implies a sum rule fixing the number (always even) and sense of the vortices attached to the particle. Also the particle drag may cause the normal fluid and superfluid vortices to couple more strongly than in the absence of particles. One can compare the drag per unit length on a vortex line, due to friction with the normal fluid[2], to the Stokes drag on a sphere on the line per sphere diameter. To the first order, this yields $3\pi\mu/\kappa\rho_s$, which is always greater than one. This suggests that the drag on the line is not influenced greatly by the presence of the particle.

The observation that particles arrange themselves with equal spacing on vortex lines is challenging to explain. Among possible explanations, we note that the de Broglie wavelength of the flow velocities is comparable to the particle spacing along the vortices. This hypothesis requires the existence of a particle-particle repulsive force that is sensitive to the de Broglie wavelength of superfluid helium.

In summary, we have directly observed vortex lines in different configurations by illuminating particles attracted to vortex cores. We have seen for the first time the structure of superfluid vortices as they are created by the phase transition, and in a steady state far from the transition temperature. The distribution and geometry of the vortices differs from predictions[2], and calls for further study and understanding. We believe that the technique offers many exciting opportunities for further work.

**Methods**  We create hydrogen particles from a room temperature mixture of hydrogen gas diluted with helium in a ratio of 1:10. We pass the mixture through a stainless steel injector of 3 mm inner diameter directly into helium I using 70 kPa of pressure in a 1 s burst. A 9.5 mm outer cylinder jackets the injector, and the space between the cylinders is sealed at both ends, allowing a cryo-pumped vacuum to develop in between. During injection, the gas cools from room temperature while moving into the cryostat. While travelling through the injector, the helium in the mixture liquefies, and the hydrogen freezes into solid particles. The mixture then flows into liquid helium within the cryostat. The particle production depends on the helium/hydrogen mixture ratio, and on the pressures at the room temperature end of the injector and within the cryostat. The hydrogen particles become smaller as the mixture ratio decreases. Most particles are below the resolution of our long-range microscope (~ 2.7 μm/pixel). Their concentration and the amount of light attenuated suggest that the particles are substantially smaller. The limited angular aperture of our cryostat precludes, for now, using light scattering experiments to estimate the particle size.

We perform experiments in a cryostat with windows on four sides. The experimental volume of helium is a 25 cm tall square cylinder with 5 cm sides. As this space is connected to a helium reservoir above, the liquid in this volume has no free surface, but remains near the liquid/vapour saturation curve. We illuminate the hydrogen particles with an argon-ion laser-sheet with an intensity of 0.5 W. The light sheet is 1.6 cm tall and ~100 μm thick. We image the plane illuminated by the sheet with two cameras: one of them being a 3000×2000 CCD array with a 7.7 μm/pixel resolution, the second being a 1000×1000 CMOS video camera with a 16 μm/pixel resolution. The entire apparatus rests on an air bearing, and the cryostat, cameras, and laser illumination can rotate at rates up to 2 Hz. The helium is cooled by evaporation with a vane-type vacuum pump. The temperature is measured using a calibrated semiconductor temperature probe.

1. Kim, E. & Chan, M.H.W. Probable observation of a supersolid helium phase. *Nature* **427**, 225-227 (2004).

2. Donnelly, R.J. *Quantized vortices in Helium II.* Cambridge University Press, 1991.

3. Onsager, L. in *Proc. Intern. Conf. Theor. Phys.,* Kyoto and Tokyo, Science Council of Japan, Tokyo, 877-880 (1953)

4. Zurek, W.H. Cosmological experiments in condensed matter systems. *Phys. Reports* **276**, 177-211 (1996).

5. Hendry, P.C., Lawson, N.S., Lee, R.A.M., McClintock, P.V.E. & Williams, C.D.H. Generation of defects in superfluid $^4$He as an analogue of the formation of cosmic strings. *Nature* **368**, 315-317 (1994).

6. Feynman, R.P. Application of quantum mechanics to liquid helium. *Progress in Low Temperature Physics I*, C.J. Gorter, ed., North-Holland Publishing Co., 17-53 (1955).

7. Vinen, W.F. Low temperature quantum turbulence: a challenge for the experimentalists. *Journal of Low Temperature Physics* **124,** 101-111 (2001)

8. Landau, L.D. & Lifshitz, E.M. *Fluid Mechanics.* Pergamon Press, 1987.

9. Koplik, J. & Levine, H. Vortex reconnection in superfluid helium. *Phys. Rev. Lett.* **71**, 1375-1378 (1993).

10. Nemirovskii, S.K. & Fiszdon, W. Chaotic quantized vortices and hydrodynamic processes in superfluid helium. *Rev. Mod. Phys.* **67**, 37-81 (1995).

11. Stalp, S.R., Niemela, J.J., Vinen, W.F. & Donnelly, R.J. Dissipation of grid turbulence in helium II. *Physics of Fluids* **14**, 1377-1379 (2002)

12. Yarmchuk, E.J., Gordon, M.J.V. & Packard, R.E. Observation of stationary vortex arrays in rotating superfluid helium. *Phys. Rev. Lett.* **43**, 214-218 (1979).

13. Batchelor, G.K. *An Introduction to Fluid Dynamics.* Cambridge University Press, 1967

14. Parks, P.E. & Donnelly, R.J. Radii of positive and negative ions in Helium II. *Phys. Rev. Lett* . **16**, 45-48 (1966).

15. Poole, D.R., Barenghi, C.F., Sergeev, Y.A. & Vinen, W.F. Motion of tracer particles in He II. *Phys. Rev. B* **71**, 064514-1-16 (2005).

**Acknowledgements** We thank Russ Donnelly, Joe Vinen, Joe Niemela, Edward Ott, Thomas Antonsen, Michael Fisher, Chris Lobb and Wojciech Zurek. This research was funded by the National Science Foundation of the U.S.A. and the Center for Superconductivity Research at the University of Maryland.

**Author Contributions** G.B. developed the injection technique and carried out the experiments. All authors discussed the results and commented on the manuscript. K.R.S. and D.P.L. contributed equally to this work.

**Author Information** The authors declare no competing financial interests. Correspondence and requests for materials should be addressed to D.P.L. (lathrop@umd.edu).